\documentclass[11pt,a4]{article}

\usepackage[dvips]{graphics}
\usepackage{epsfig}

\newcounter{ctr}
\renewcommand{\thectr}{(\roman{ctr})}

\begin{document}
\title{Theory of Robustness of Irreversible Differentiation in a Stem Cell System: Chaos hypothesis}
\author{ Chikara Furusawa and
        Kunihiko Kaneko \\
        {\small \sl Department of Pure and Applied Sciences}\\
        {\small \sl University of Tokyo, Komaba, Meguro-ku, Tokyo 153, JAPAN}
\\}
\date{}
\maketitle
\begin{abstract}
Based on extensive study of a dynamical systems model of the development of a cell society, 
a novel theory for stem cell differentiation and its regulation is proposed
as the ``chaos hypothesis''.
Two fundamental features of stem cell systems - stochastic differentiation of stem 
cells and the robustness of a system due to regulation of this differentiation - are found
to be general properties of a system of interacting cells exhibiting chaotic intra-cellular reaction 
dynamics and cell division, whose presence does not depend on the detail of the model.    
It is found that stem cells differentiate into other cell types stochastically due to a dynamical instability 
caused by cell-cell interactions, in a manner described by the Isologous Diversification theory.        
This developmental process is shown to be stable not only with respect to molecular fluctuations 
but also with respect to removal of cells.
With this developmental process, the irreversible loss of multipotency accompanying the change from a stem cell
to a differentiated cell is shown to be characterized by a decrease in the chemical
diversity in the cell and of the complexity of the cellular dynamics.          
The relationship between the division speed and this loss of multipotency is also discussed.
Using our model, some predictions that can be tested experimentally are made for a stem cell system. 
\end{abstract}

\section{Introduction: Some Biological Facts regarding Stem Cells}

Stem cells are defined as cells that are capable of proliferation, self-maintenance of their population, 
production of differentiated cells, and regeneration of the tissue in concern \cite{Potten}.
In mammals, stem cells sustain tissues composed of cells that have a limited life span and lose 
the ability of self-renewal (e.g. blood and epidermis cells), by continuously supplying
new differentiated cells.
The importance of stem cell systems, however, is not restricted only to maintaining such kinds of tissues.
For example, in primitive multicellular organisms such as hydra \cite{hydra} and planarian \cite{planarian}, 
the regulated proliferation and differentiation of stem cells both serves to replace
differentiated cells that die through the normal course of aging and allows for the regeneration of parts of the organisms damaged through injury and the asexual reproduction of the organism itself.
In plants, development and regeneration are also sustained by stem cells at the apical meristem \cite{cell}.
These facts suggest that the function of stem cells can be regarded as fundamental in creating and maintaining cellular diversity, and that they play a most important role in sustaining the reproduction process of multicellular organisms, including development, regeneration and ordinary replacement of differentiated cells.

An important feature of a stem cell system is the 
robustness of the system with regard to cell population, 
as exemplified by regeneration.  When tissue is
damaged, active stem cells reappear through the activation of quiescent stem cells 
or the de-differentiation of neighboring cells.
The proliferation and differentiation of these stem cells is regulated, depending on the environment in such a way that the damaged tissue can be repaired.
For example, in a hematopoietic system, when the number of terminal 
differentiated cells of a particular type (e.g. red blood cells) is decreased due to some externally applied influence, the production of this cell type through differentiation is increased, and the original population distribution is recovered \cite{cell}.

Although the behavior of stem cells has been studied intensely because of their
biological and medical importance, the problem of how stem cells determine their own fates (either self-renewal or differentiation) is still unsolved.
Concerning the ``decision" of a stem cell to differentiate, 
two major hypotheses 
have been proposed.  One of these is the cell-intrinsic hypothesis, according to which 
stem cell differentiation is triggered by an intra-cellular program.
The other is the inductive hypothesis, according to which differentiation is 
caused by inductive signals received from outside the system, which cause a change in some stem cells that leads to differentiation.

A most important model representing the cell-intrinsic hypothesis is the stochastic 
differentiation model, proposed to describe the hematopoietic stem cell system by Till and 
McCulloch \cite{Till}. They observed heterogeneity in the number of stem cells in 
spleen colonies which originate from a single stem cell.
They explained this heterogeneity in terms of the hypothesis that each decision of a stem cell to either undergo self-renew or differentiation is stochastic, with a given fixed probability.
Nakahata et al. extended this stochastic determination hypothesis to the 
commitment of differentiation from multipotential stem cells in a hematopoietic system \cite{Nakahata}.
They prepared two daughter cells derived from a single parent stem cell and 
cultivated them independently under identical conditions.
Repeating this experiment a number of times, they found that the distributions of two such colonies generated by paired daughter cells often turned out to be different from each other.
These experiments clearly demonstrate that microscopic fluctuations of 
an intra-cellular state can be amplified, and lead to stochastic behavior of stem cells.
Following these experiments, the stochastic differentiation model is extended to other stem cell systems, such as those of the central nervous system \cite{neuro_stochastic}, hydra \cite{hydra_stochastic}, and 
interstitial crypts \cite{interstitial_stochastic}.

In experiments on several stem cell systems, spontaneous differentiation 
has been observed, and in terms of the ovservables considered in those studies,
it appears that it occurs in a stochastic manner.
Although external signal molecules, such as growth factors, are required to maintain 
these systems, it has been found that the differentiation of stem cells does not require any signals from the outside providing inhomogeneity among the state of stem cells to govern their decision to differentiate or not.
From these results, it can be concluded that the inductive hypothesis of stem cell differentiation is incorrect, and that diversity comes from inside the system.

It should be noted that the cell-intrinsic hypothesis does not exclude the
possibility  that extrinsic factors can affect a stem cell's behavior.
Certainly, signal molecules can regulate the behavior of cells in 
a stem cell system.
For example, experiments using hematopoietic stem cells both in vivo and in vitro 
suggest that the presence of signal molecules, such as CSFs(colony stimulating 
factors) and interleukins, is essential for the survival and proliferation 
of the cells \cite{Ogawa}.
In vivo, these signal molecules, including diffusive chemicals and cell 
surface proteins, are released by cells of several types, such as stem cells, 
differentiated cells, and neighboring cells, like stromal cells in bone marrow.
These signals sustain the complex interactions among these cells in the stem 
cell system.
These complex interaction are responsible for the robustness of such systems that is generally observed.

Summarizing this discussion, it is now clear that both stochastic differentiation and complex cell-cell 
interactions are essential in stem cell systems.
However, an interesting and important question remaining here is whether the nature of this differentiation depends on these cell-cell interactions.
In other words, it is yet unknown whether the probabilities involved in differentiation are controlled extrinsically.

The original form of the stochastic model assumes that the probability of differentiation 
is fixed, and that cell population is regulated by controlling 
the proliferation and survival of cells.
However, there are some experiments which suggest that this probability can be 
controlled by extrinsic factors.

For example, when a single hematopoietic stem cell is transplanted into a mouse whose hematopoietic system has been destroyed through irradiation,
this cell can reconstitute the hematopoietic system with high probability \cite{Osawa}.
This result implies that, the probability of differentiation is externally controlled at least in the case that there are only a small number of stem cells.
The regulation of this probability has also been suggested in the spontaneous differentiation of hydra stem cells \cite{hydra_stochastic}.

Indeed, regulation of the probability of differentiation is required for a developmental process that is stable with respect to some external perturbations.
For example, let us consider differentiation from
a stem cell to some other cell types. Assume that
cells of one of these types is continually eliminated.
Then, if there were no regulation of the rate of differentiation from the
stem cell to the various cell types, 
the population of the types being eliminated would gradually decrease, with potentially tragic results.
Even if the speed of overall cell division of the stem cell
were increased with no adjustment regarding cell types, the original distribution of cell types could not be recovered,
since in this case, maintaining the population of the damaged type would imply increased population of the other types.
If the rate of differentiation to a terminal cell type is $F$, and the death rate is $\gamma$,
the population of this terminal cell type in a steady state is given by $N = F/\gamma$.
(This is obtained by solving for the steady state of the equation describing the cell-population dynamics, $dN/dt=F-\gamma N$.
Even if there exist several intermediate 
stages of differentiation from the stem cell to the terminal cell, this conclusion is true for the steady state.)
Thus the flow $F$ to this branch from the stem cell should increase when 
the death rate $\gamma$ of the terminal cell is increased.  

Thus there obviously exists some type of regulation of
differentiation, and, as discussed above, there is experimental
evidence suggesting that this regulation is provided externally.

\section{Requirement to Understand the Logic of Stem Cell}

Summarizing the discussion of the previous section, the following features characterize a stem cell system.

\begin{enumerate}

\item

The diversity of cell types resulting from differentiation from a stem cell  
emerges without any influence from the outside of the system.
This differentiation occurs stochastically.
 
\item
The proliferation and differentiation of cells in a system is regulated 
by cell-cell interactions.
This regulation keeps the system stable with respect to external perturbations. 

\end{enumerate}

In the present paper, we show that these features are a 
general consequence of intra-cellular chemical dynamics and
cell-cell interactions, and their existence does not depend on the details of the specific chemical substances present.
In particular, stochastic differentiation and the regulation of the 
probabilities involved in differentiation arise naturally as general properties
of interacting dynamical systems.

Note that the origin of stochastic differentiation is yet unknown in the context of molecular biology.
A simple explanation for this could be random 
collisions of specific chemical substances in  a cell.
For example, such stochasiticy is expected in the case that 
differentiation is caused by the reaction of a DNA molecule and a regulator protein through a chance encounter.

If such uncontrollable random collisions were the sole cause of stochastic differentiation,  however, it would be very difficult
to imagine a mechanism to regulate the probability itself.
If differentiation were caused by a specific stochastic reaction of this kind, 
this reaction should be controlled by other stochastic chemical reactions.
These chemical reactions may form a complex reaction network, in which each reaction 
often is accompanied by a large fluctuation arising from stochastic 
reactions among a small number of chemical substances.
Now, exercising control to change the rate of some specific reaction in such a network 
would probably be very difficult.

In the present paper, we do not propose a sophisticated
control mechanism capable of exercising precise control over such a specific reaction.
Rather, we show that  stochastic differentiation with autonomous regulation can emerge from a complex reaction network in general, without any fine-tuned mechanisms.

Here, we note that chaotic dynamics provides a plausible mechanism
for such control of the probabilistic behavior involved in differentiation, because a set of deterministic
equations that describe biochemical reaction networks in a cell can yield stochastic behavior in the form of chaotic dynamics. 
As long as information concerning the  internal 
cellular state is not precisely known to us,
differentiation must appear stochastic. 
On the other hand, since this ``stochastic" behavior is generated by a deterministic mechanism, 
it is possible that chaotic dynamics of the biochemical reaction networks in cells provide
a mechanism for the self-regulation of differentiation.

In the present paper, we propose this `{\bf chaos hypothesis}' of stem cell differentiation, in which the nonlinear dynamics of intra-cellular chemical reactions and 
cell-cell interactions are essential. Indeed,
the authors and T. Yomo have carried out several numerical experiments modeling interacting cells 
with internal biochemical networks and cell divisions, and proposed
the Isologous Diversification theory as a general mechanism of 
the spontaneous differentiation of replicating biological units \cite{KKTY1, KKTY2, CFKK1, CFKK2, KKTY3}.
The following three points of this theory are relevant to the consideration of stem cell systems.

\begin{enumerate}

\item 
Spontaneous differentiation:

Cells exhibiting oscillatory chemical reactions 
differentiate through interaction with other cells.
Cells first split into groups with different phases of oscillation, 
as described by the theory of
clustering in coupled dynamical systems \cite{GCM1,GCM2}, 
and then they split into groups with different chemical compositions.
This differentiation results from the generation of new states
corresponding to distinct orbits separated within the phase space of the set of equations used to describe the chemical reactions within a cell.
Each differentiated state of a cell starts to be preserved by the cell division
and is transmitted to its offspring.
These differentiations are not caused by specific substances, but, rather, are 
brought about by orbital instabilities inherent in the underlying nonlinear dynamics.

\item 
Emergence of a hierarchical set of differentiation rules:

These differentiations obey specific rules, which emerges form inter- and intra- dynamics.
For example, pluripotent cells, like stem cells, have the potentiality into 
differentiate into committed cells, which can further differentiate to terminally differentiated cells.
Such differentiation processes often are stochastic, and the probability for 
each outcome depends on the distribution of cell types in the system.

\item
Stability of the distribution of cell types:

As a result of regulated differentiation, the population dynamics of 
the cells of each type begin to follow a rule at a higher level than
that governing chemical concentrations.
In general, the relative number of each cell type approaches a steady state by 
this population dynamics has a stable attractor.
This implies that the resulting distribution of cell types is stable with respect to 
external perturbations up to a certain degree.
For instance, when the population of one type of cell is decreased due to some external influence, 
the original distribution is recovered through the increase of differentiation to this cell type.
In this theory, although the instability of the dynamical system triggers differentiation, 
the cell society as a whole is stabilized through cell-to-cell interactions.

\end{enumerate}

In this paper, we attempt to capture characteristic dynamics of stem cell systems using the above theoretical framework.
The {\bf chaos hypothesis} for stem cell dynamics is proposed, according to which
stochastic differentiation is a result of chaotic dynamics of the cell system
arising from intra-cellular chemical reactions and cell-cell interactions.
These differentiations are controlled in a manner that depends on the distribution of cell types.
This control results in a system whose overall features are stable even without the presence of an external regulation mechanism.
Irreversible differentiation leading to a loss of multipotency is characterized theoretically.


\section{Model}

First, we summarize our standpoint for designing the model of 
a stem cell system.

Recent experimental studies of stem cell systems have shown that intra-cellular dynamics and cell-cell interactions involve many kinds of chemical species and that the interactions among these chemicals form a complex network.
Due to the complexity of cellular dynamics, it is almost
impossible to construct a model that gives results in precise agreement with the quantitative behavior of real biological processes.
Of course, one could go about constructing a complicated model with sole purpose of precisely reproducing experimental data obtained in cell biology studies and in this way imitate the behavior of living cells.
However, such mimicry would get us no closer to understanding the essence of cellular systems, because any similar behavior to real biological processes could be obtained by adding complicated mechanisms.
With such a model, we can neither identify the essential features of stem cells nor extract universal features that are present in all stem cell systems.
By contrast, we start with a simple model containing only the essential features of biological systems, and with this we attempt to capture the universal behavior exhibited by all cell societies possessing such features, such as cellular differentiation from a multipotent stem cell, and irreversible determination of differentiation.

To investigate cell differentiation and the robustness of cell societies, the authors and T. Yomo have considered simple models consisting only of basic cell features.
We have found that these simple models possess the essential features of stem cell systems - stochastic differentiation from a stem cell and 
robust development of a cell society through regulation of the differentiations - as long as the intra-cellular dynamics are sufficiently complex 
to exhibit non-linear oscillatory behavior including chaos.

Our model consists of the following four parts:

\begin{itemize}

\item { Internal dynamics consisting of a biochemical reaction network within each cell.}

\item { Interactions between cells (inter-cellular  dynamics).}

\item { Cell division and cell death. }

\item { Molecular fluctuations in the intra-cellular reaction dynamics. }

\end{itemize}

In essence, we assume a network of catalytic reactions for the internal dynamics that 
displays non-linear oscillations of chemicals, and a cell-cell interaction process 
that consists simply of diffusion of chemicals through media.
Below we describe each process.

\subsection{Internal chemical reaction dynamics}

Within each cell, there is a network of biochemical reactions.
This network includes not only a complicated metabolic network 
but also reactions associated with genetic expressions, signaling pathways, 
and so forth.  In the present model, a cellular state is represented by
the concentrations of $k$ chemicals.
The dynamics of the internal state of each cell is expressed by a set of variables 
$ \{ x^{(1)} _i(t),\cdots,x^{(k)} _i(t) \} $ representing the concentrations of
the $k$ chemical species in the {\it i}-th cell at time {\it t}.

As the internal chemical reaction dynamics, we choose a catalytic network among the $k$ chemicals.
Each reaction from some chemical $i$ to some chemical $j$ is assumed to be catalyzed by a third chemical $\ell$, which is determined randomly. 
To represent the reaction-matrix we use the notation\footnote{Precisely speaking, this should be called tensor.} $Con(i,j,\ell)$, which takes the value 1 
when the reaction from chemical $i$ to chemical $j$ is catalyzed by $\ell$, and 0 otherwise.
Each chemical acts as a substrate to create several enzymes for other reactions and there are several paths to other chemicals from each chemical.
Thus these reactions form a complicated network.
The matrix $Con(i,j,\ell)$ is generated randomly before a simulation, and is fixed throughout this simulation.

We denote that the rate of increase of $x^{(m)}_i(t)$ (and decrease of $x^{(j)}_i(t)$) through a reaction from chemical $m$ to chemical $j$ catalyzed by $\ell$ as $e x^{(j)}_i(t) (x^{(\ell)}_i(t))^{\alpha}$, where $e$ is the coefficient for the chemical reaction and $\alpha$ is the degree of catalyzation.
For simplicity, we use identical values of $e$ and $\alpha$ for all paths.
In this paper, we adopt $\alpha = 2$, which implies a quadratic effect of enzymes.
This specific choice of a quadratic effect is not essential in our model of cell differentiation.
There is a large class of forms for the internal dynamics that result in qualitatively similar behavior.
The important point here is only the existence of nonlinear oscillation.

Furthermore, we take into account the change in the volume of a cell, which varies as a result of the transportation of chemicals between the cell and the environment.  
For simplicity, we assume that the total concentration of chemicals
in a cell is constant, $\sum_m x^{(m)}_i = const$. It follows that the volume of a cell is proportional to the sum of the quantities of all chemicals in the cell. The volume change is calculated from the transport, as discussed below.

\subsection{Cell-cell interaction}

Cells interact with each other through the transport of chemicals into and out of 
the surrounding medium.
The ``medium" here is not mean external to the organism, but, rather, it refers to the 
interstitial environment of each cell.
Here, the state of the medium is expressed by a set of variables $ \{ X^{(1)} (t),\cdots,X^{(k)} (t) \} $, whose elements represent the concentrations of the $k$ chemical species in the medium.

We assume that the medium is well stirred by ignoring the spatial variation of the concentration, so that all cells interact with each other through an identical environment.
Of course, this is a simplifying assumption that introduces a limitation of our model, since spatial factors, such as the gradient of signal molecules, play an essential role in morphogenesis.
However, experiments in vitro show that a diversity of cell types in stem cell systems can appear even in a homogeneous environment, without any spatial structures.
Thus the appearance of a diversity of cell types does not require some kind of spatial inhomogeneity, which is often introduced from outside the system in the modeling of stem cells behavior.
Instead, the diversity of cell types emerges through processes within the system.
In fact, we have also investigated a model possessing spatial structure, to study pattern formation.  
However, even in this case, the differentiation is not due to this spatial pattern, but rather to the interactions between cells. The spatial pattern is formed later as a result of the differentiation.
The cells in our simulations are in general grown in a homogeneous environment, where there is no spatial variation of chemical concentrations.

In this model, we consider only indirect cell--cell interactions in the form of the diffusion of the chemical substances in the system, as a minimal form of interaction.
We assume that the rate at which a chemical is transported into a cell is proportional to the difference between the concentration of this chemical inside and outside the cell.
Thus, the term in the equation for each  $d x_i^{(m)}(t)  /dt$ describing the transport from 
the medium into the $i$-th cell for $m$-th chemical is given by $D(X^{(m)} (t) - x^{(m)}_i (t))$
, where $D$ is a transport coefficient.
Of course, the transport through a membrane is not this simple, involving several mechanisms such as endocytosis and transport through channel proteins.
However, the specific form of this interaction is not important to describe the general behavior in which we are interested, and therefore we ignore these complicated processes for simplicity.

In general, the transport (diffusion) coefficients should be different for different chemicals.  
Here, we consider a simple situation in which there are two types of chemicals, those which can penetrate the membrane and those which cannot.
To account for this, we use two values for the diffusion coefficient, $D$ in the former case and 0 in th latter.
This distinction is made in the equation of motion by the parameter $\sigma_m$, which is 1 if the $m$-th chemical is penetrating and 0 if it is not.

\subsection{Cell division and cell death}

Each cell receives penetrating chemicals from the medium as nutrients,
while the reaction in the cell transforms them into non-penetrating 
chemicals which comprise the body of the cell.
As a result of these reactions, the volume of the cell is
increased.
In this model, the cell divides into two almost
identical cells when the volume of the cell is doubled.

We assume that two cells arising from a single cell contain almost identical chemical
compositions.
``Almost" here means that the concentrations of chemicals in a daughter cell are slightly different from the mother's. The chemical concentrations of divided cells are $(1+\epsilon)x^{(m)}$ and $(1-\epsilon)x^{(m)}$ respectively with a small ``noise" $ \epsilon $, a random number with a small amplitude, say over $[-10^{-6}, 10^{-6}]$.
In our model simulations, this tiny difference can be amplified and differentiation occurs among the cells.
The important feature of our model is the amplification of microscopic differences between the cells resulting from an intrinsic instability of the internal dynamics.

Penetrating chemicals can penetrate the cell membrane in both directions, and thus the diffusive nature of the system implies that in certain cases, these chemicals will flow out of a cell.
As a result, the volume of the cell becomes smaller.
In our model, a cell dies when its cell volume become less than a given threshold.

\subsection{Molecular fluctuations}

In a real biological system, the number of a specific type of molecules in a cell is often small. For example, some signal molecules can cause a change of the cellular state even when the number of molecules is the order of 1000.
With numbers of molecules on this order, it follows that the cellular dynamics always exhibit non-negligible molecular fluctuations.
Thus a noise term should be included to take into account (thermal) fluctuations.
Considering fluctuations of $\sqrt{N}$ for a reaction involving $N$ molecules, we add a noise term proportional to $ \eta(t) \sqrt{x^{(m)}_i(t)}$, corresponding to a Langevin equation, where $\eta(t)$ denotes Gaussian white noise satisfying \\
$ <\eta^{(m)}_i (t) \eta^{(m')}_i (t')>= \sigma ^2 \delta(t-t')$, with $\sigma$ the amplitude of the noise.
~\\

Summing up all of the processes described above, the dynamics of the chemical concentration in each cell is represented as follows:

\begin{equation}
dx^{(\ell)}_i(t)/dt  =  \delta x^{(\ell)}_i(t) - x^{(\ell)}_i(t) \sum_{m=1}^k\delta x^{(m)}_i(t), 
\end{equation}
with

\begin{eqnarray}
\delta x^{(\ell)}_i(t)  & =  & ~~~\sum_{m,j}Con(m,{\ell},j) \;e \;x^{(m)}_i(t) \;(x^{(j)}_i(t))^{\alpha} \nonumber \\
& & - \sum_{m',j'}Con({\ell},m',j') \;e \;x^{({\ell})}_i(t) \;(x^{(j')}_i(t))^{\alpha}  \nonumber \\
& & + \sigma_{\ell} D(X^{(\ell)} (t) -x^{({\ell})}_i (t)) \nonumber \\
& & + \eta(t) \sqrt{x^{(\ell)}_i(t)}.
\end{eqnarray}

\noindent
Here, the terms with $\sum_{m,j}Con(m,{\ell},j)$ and $\sum_{m',j'}Con({\ell},m',j')$ represents paths going into $\ell$ and out of $\ell$, respectively.
The term $\delta x_i^{(\ell)}$ is the increment of chemical $\ell$, while the second term in 
Eq.(1) introduces the constraint of $\sum_{\ell}x^{(\ell)}_i(t)=1$ due to the growth of the volume.
The third term in Eq.(2) represents the transport between the medium and the cell, and the fourth term represents the molecular fluctuations.

Since the penetrating chemicals in the medium are consumed through their flow into the cells, we must supply these chemicals (nutrition) into the medium from the outside.
Denoting the external concentration of the $\ell$-th such chemical by $\overline{X^{(\ell)}}$ and the transport coefficient of the flow by $f$, its dynamics are described by 

\begin{equation}
dX^{(\ell)}(t)/dt  =f\sigma_{\ell}(\overline{X^{(\ell)}} -X^{(\ell)}(t))-(1/V)\sum_{i=1}^N { \sigma_{\ell}D(X^{(\ell)}(t)-x^{(\ell)}_i(t)) }, 
\end{equation}
where {\it N} denotes the number of the cells in the environment, and $V$ denotes the volume of the medium in units of the initial cell size.

\subsection{Requisite for intra-cellular dynamics}

\begin{figure}[htbp]
\begin{center}
\includegraphics[width=15cm,height=12cm]{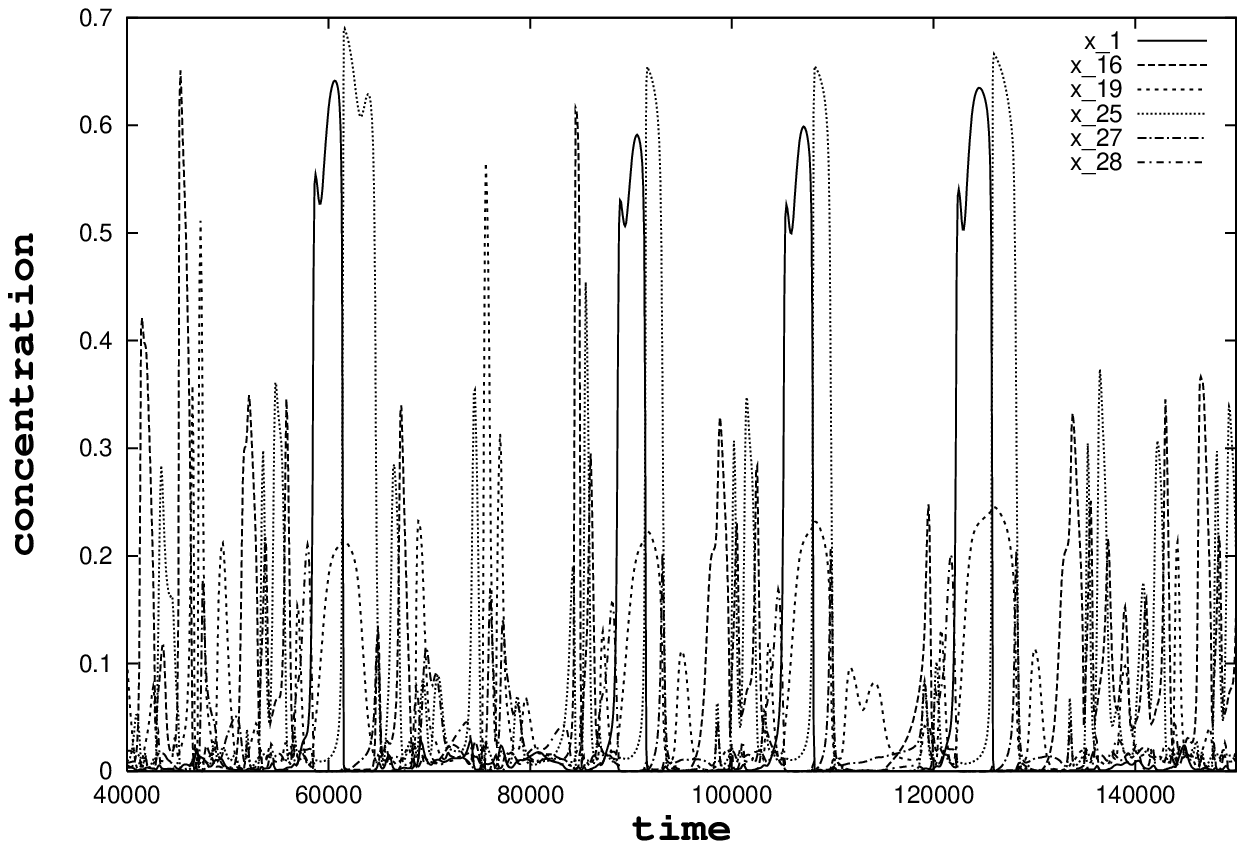}
\end{center}
\caption{\label{type-0}
Overlaid time series of $x^{(m)}(t)$ for the type-0 cell, obtained from a randomly generated network with 32 chemicals and 9 connections for each chemical.
The vertical axis represents the concentration of chemicals and the horizontal axis represents time.
In this figure, we have plotted the time series of only 6 of the 32 internal chemicals, for clarity.
The lines designated by the numbers $m$=1,16,19,25,27, and 28 represent the time series of the concentrations of the corresponding chemicals $x^{(m)}(t)$.
The parameters ware set as $e$=1.0, $D$=0.001, $f$=0.02, $\overline{X^{(\ell)}}=0.2$ for all $\ell$, and $V=10$.
Chemicals $x^{(m)}(t)$ for $m<10$ are penetrating (i.e., $\sigma_{m}=1$), and others are not. 
}
\end{figure}

The dynamics of the chemical concentrations depend on the choice
of the reaction network.  
They can be simple, like fixed-point or limit cycle dynamics, or they can be complex, like chaotic dynamics.
Here we assume that the intra-cellular dynamics 
for a stem cell includes some dynamical instability, which typically leads to oscillatory behavior including chaos, as depicted in Fig.\ref{type-0}.

In our model simulations, when the number of paths in the reaction matrix is small, the cellular dynamics generally fall into a steady state without oscillation, in which a small number of chemicals are dominant, while other chemicals' concentrations vanish.
On the other hand, when the number of reaction paths is large, all chemicals are generated by other chemicals, and chemical concentration come to realize constant value (which are almost equal for many chemicals).
Only when there is an intermediate number of reaction paths do non-trivial oscillations of chemicals appear as seen in Fig.\ref{type-0}. 
Another factor relevant to the nature of the internal dynamics is the number of auto-catalytic paths.
We find that the probability of the appearance of oscillatory dynamics increases as the number of auto-catalytic paths increases.
It is important to note, however,  that even by using appropriate parameter values the 
cellular dynamics resulting from randomly chosen reaction networks are rarely oscillatory.
We have performed simulations with our model using thousands of different reaction networks, and only about 10\% of these networks resulted in oscillatory dynamics.

The reasons for studying networks that display oscillatory dynamics are as follows.

First, as shown throughout this paper, we have found that robust developmental processes with spontaneous differentiation of stem-type cells  emerges commonly if and only if the individual cells exhibit oscillatory dynamics.
Second, as discussed in \S\ref{Discussion} and in Ref.\cite{PRL}, a cell system
characterized by oscillatory dynamics has a higher growth speed as an ensemble,
and for this reason it is expected to be selected through evolution.

In real biological systems, oscillations are observed in some chemical substrates such as $Ca^{2+}$, NADH, cyclic AMP, and cyclins.
These oscillations are sustained by networks of non-linear chemical reactions, often including positive feedback reactions.
Experimental results show that in stem cell systems, reactions among signal molecules form a complex network with positive and negative feedback \cite{Ogawa}.
Thus, it is natural to postulate the existence of such oscillatory dynamics in our model system.
The importance of oscillatory dynamics in cellular systems has been pointed out by Goodwin \cite{Goodwin}.

\section{Numerical results}

In this section, we present numerical results demonstrating development with stem-cell-type behaviors obtained from several simulations.
Here we consider numerical experiments employing a particular reaction network with the number of chemicals $k$ fixed to 32.
This case represents typical behavior of the developmental process.
The results presented in this section were obtained without the use of molecular fluctuations.
As we discuss in \S\ref{sta_s}, however, they are stable with respect to noise up to a particular threshold strength.

\subsection{Differentiation process}

As the initial state, we consider a single cell whose chemical concentrations $x^{(\ell)}_i$ are determined randomly in such a way to satisfy the constraint $\sum_{\ell} x^{(\ell)}_i  =1$.
In Fig.\ref{type-0}, we have plotted a time series of the concentrations of the chemicals in such a single, isolated cell.
The attractor of the internal chemical dynamics depicted here is chaotic.
Here, most of the chemicals are generated by other chemicals and coexist within the cell.
We call cells in such a state ``type-0'' in this paper.
This state appears dominantly with randomly chosen initial conditions for a single, isolated cell; in other words, the basin of attraction for this state is very large\footnote{
In the case of single-cell dynamics considered here, there are very rare
initial conditions that results in a fixed-point attractor, as shown below.
}.

Now, considering the case with diffusion, external chemicals flow into the
cell, because of lower concentration of penetrating chemicals in the cell by reactions transforming penetrating chemicals into non-penetrating chemicals.
This flow leads to the increase of the cell volume.
If this volume becomes sufficiently large, the cell divides into two, with almost identical chemical concentrations.
The chemicals within the two daughter cells oscillate coherently, with the same
dynamical behavior as their mother cell (i.e., type-0 dynamics).  
As the number of cells increases by a factor of two (i.e., $1 \rightarrow 2 \rightarrow 4  \rightarrow 8 \cdots$)
with further divisions, although the internal dynamics of each cell correspond to the same attractor, the coherence of the oscillations among individual cells is easily lost due to the chaotic nature of the dynamics in these type-0 cells.
The microscopic differences introduced at each cell division are amplified to a macroscopic level, and this destroys the phase coherence.

\begin{figure}[htbp]
\begin{center}
\includegraphics[width=15cm,height=19cm]{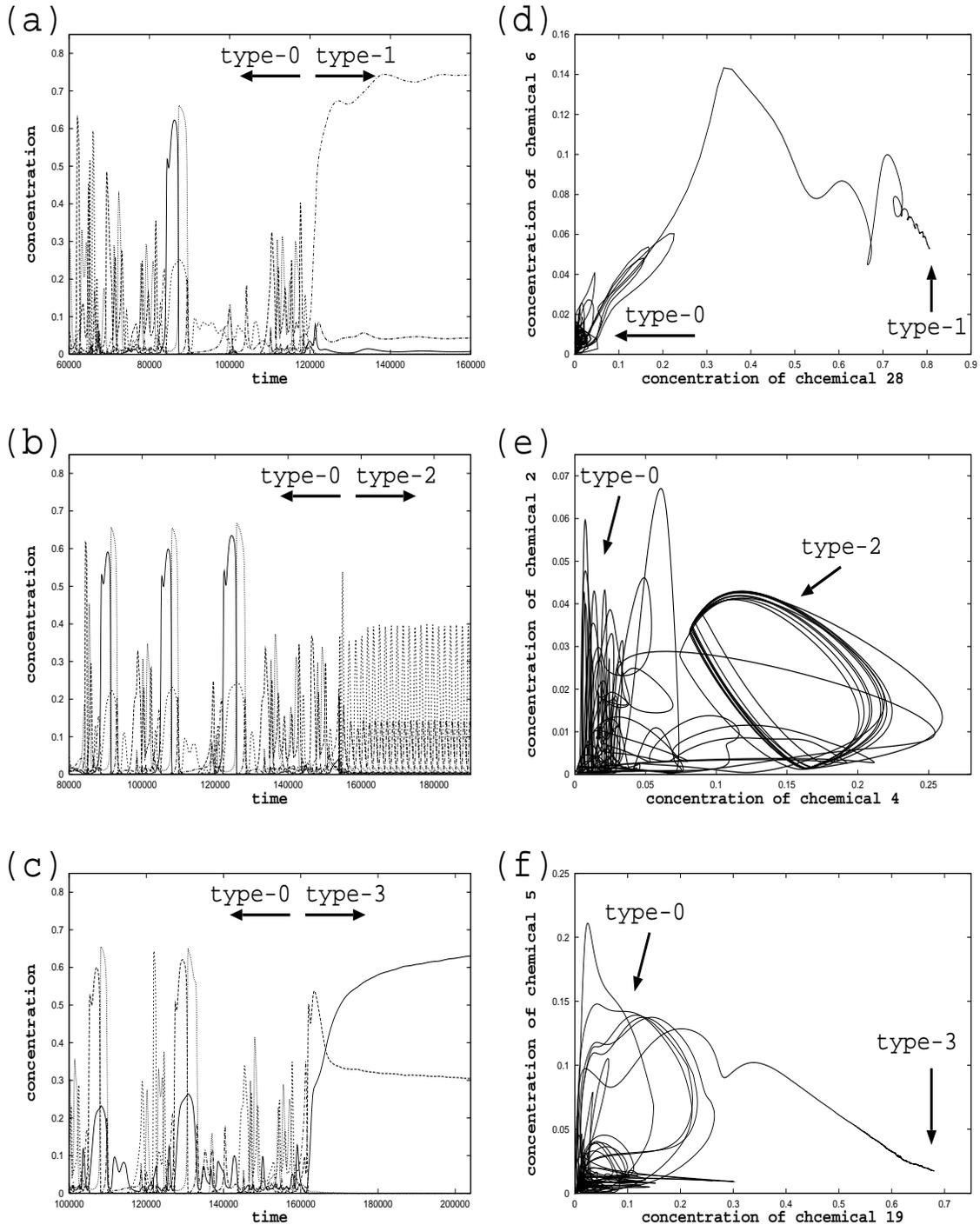}
\end{center}
\caption{\label{diff}
(a)-(c): Time series of $x^{(m)}(t)$, overlaid for 6 chemicals (as in Fig.\ref{type-0}) in a cell, representing the course of differentiation to type-1,2, and type-3 cells, respectively.
(d)-(f): Orbit of the internal chemical dynamics in the phase space projected onto $(x^{(\ell)},x^{(m)})$, with $(\ell,m) =$(28,6) in (d), (4,2) in (e), and (29,5) in (f), corresponding to (a)-(c).
The orbit of the chemical concentrations at a transient process from type-0 to type-1, 2, and type-3 cells are plotted in the projected space.
Note that we consider different chemicals in each figure, to make each cell type clearly distinguishable.
The orbits of type-1 and type-3 cells fall onto fixed points (see (d) and (f)), while the orbit of the type-2 cell falls into a limit-cycle attractor (see (e)).
}
\end{figure}

When the number of cells exceeds some threshold value, some cells 
begin to display different types of dynamics.
The threshold number depends on the nature of the networks and the parameter values used in our model.
In the situation we consider presently, two types of cells with distinct internal dynamics begin to emerge from type-0 cells when the total cell number becomes 64.
In Figs.\ref{diff}(a) and (b), the time series of the chemical concentrations in these new types of cells are plotted.
We call these ``type-1" and ``type-2" cells, respectively.
In Figs.\ref{diff}(d) and (e), orbits of the chemical concentrations at the transition from type-0 cells to types 1 and 2 are plotted in the phase space.
These figures show that each of the these attractors occupies a distinct region in the phase space, and that each can be clearly distinguished as a distinct state.
These transitions of the cellular state are interpreted as differentiation.

Note that this differentiation is not induced directly by the tiny differences introduced at the cell divisions.  
Also, the transition from one cell type to another does not occur at the time of 
cell division, rather later through the interaction among the cells.  
The phenomenon of differentiation observed here is caused by an instability in the entire dynamical system consisting of all the cells and the medium.  
It is due to this instability that tiny differences between two daughter cells can be come amplified to a macroscopic level through the intra-cellular dynamics and interactions between cells.
Only when the strength of this instability exceeds the threshold, does differentiation occur.  
Then, the emergence of new cell types stabilizes the dynamics of the other cell types.  
The cell differentiation process in our model is due to the amplification of tiny phase differences through orbital instability (transient chaos), while the coexistence of different cell types stabilizes the system as a whole.

The important point here is that the spontaneous differentiation from the type-0 cells is ``stochastic".
The choice of a type-0 cell to either replicate or differentiate
appears stochastic if we consider only the cell type.
Of course, if one could determine completely the
chemical composition and the division-induced asymmetry, then
the fate of the cell would also be determined, since our model 
is deterministic.  
Here, however, 
it would be practically impossible to make such a determination because it would necessitate of the internal cell state to arbitrary precision, given the chaotic nature of its dynamics.

\begin{figure}[htbp]
\begin{center}
\includegraphics[width=13cm,height=13cm]{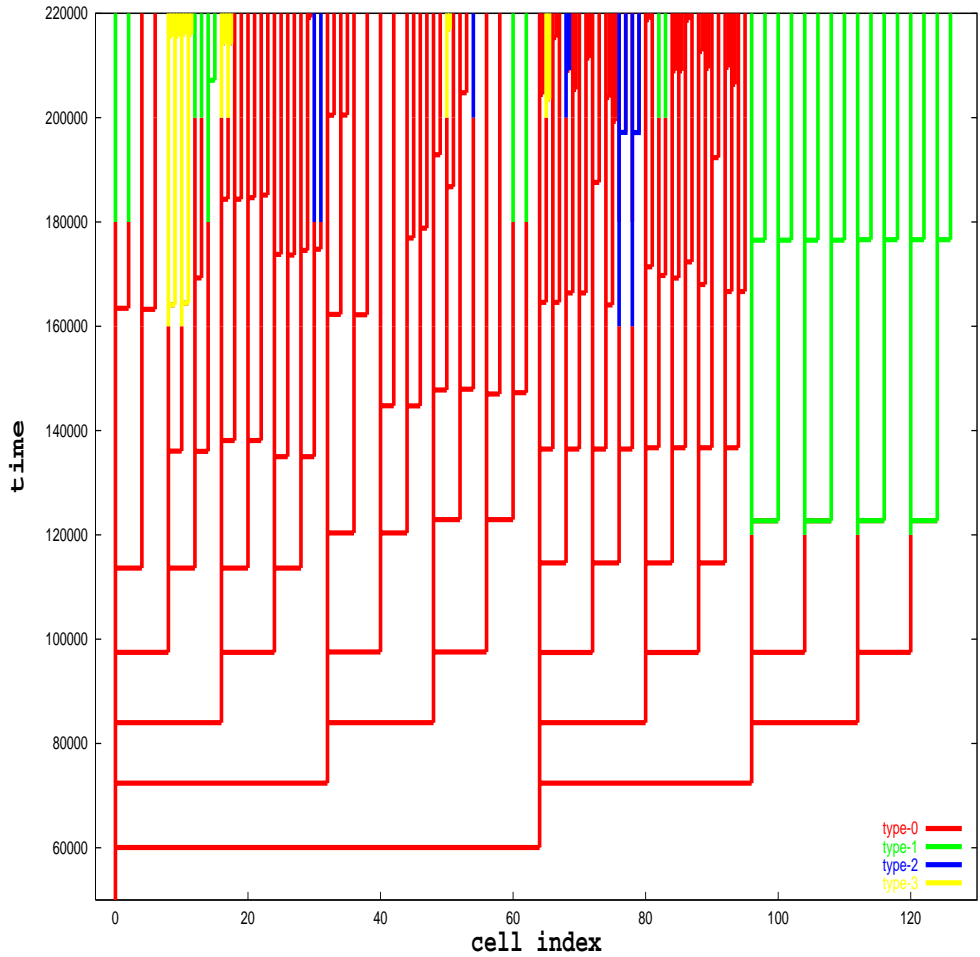}
\end{center}
\caption{\label{lineage}
Cell lineage diagram.
The vertical axis represents time, while the horizontal axis represents a cell index, which is 
defined only for the practical purpose of avoiding line crossing.
In this diagram, each bifurcation of lines through the horizontal segments
corresponds to the division of the cell, while each vertical line represents an existing cell, with the color indicating its cell type.
}
\end{figure}

As the cell number increases further, another type of cell appears through differentiation from type-0 cells, which we call ``type-3" cells (see Figs.\ref{diff}(c) and (f)).
In this example, eventually four types of cells appear through the developmental process.
At this stage, the differentiation is determined, and a sort of memory is formed in the cellular dynamics, as was first discussed by one of the present authors (KK) and T. Yomo \cite{KKTY1, KKTY2}.
We can thus draw the cell lineage diagram, as shown in Fig.\ref{lineage}, where a division process is represented by the connection between the lines corresponding to mother and 
daughter cells, while the color of a line indicates the cell type.

It should be noted that these differentiated cell types do not necessarily correspond to attractors of the internal dynamics of a single cell.
For the interaction matrix and parameter values used in the simulation described above,
for a single cell system, there is no attractor corresponding to type-2 dynamics.
In a multiple-cell system, the stability of such internal dynamics is sustained only through interactions among cells.
This means that when a single type-2 cell is transplanted into a medium containing no other cells, this cell is de-differentiated back into a type-0 cell.
On the other hand, type-1 and type-3 dynamics do exist as attractors of the single-cell system, like type-0 dynamics.
However, these two attractors are not observed when the initial conditions (i.e. the initial concentrations of chemicals in the first cell) are chosen randomly.
This reveals that the basins of attraction for these states are much smaller than that of the type-0 state.

\subsection{Emergence of rules for differentiation}

The transitions between different cell types through differentiation follows specific rules.
These rules originate in a constraint on the transient dynamics exhibited between 
the attractor states corresponding to each cell type. 
In Fig.\ref{auto}, we present an automaton-like representation of these rules. 
The node `0' has four paths, one to itself, and the others to the nodes `1', `2', and `3', while each of the nodes `1', `2', and `3' has only one path, which leads back to itself.
Paths leading from one cell type back to the same type represent replication of the same cell type through division, while the other paths represent differentiation into the corresponding cell types.
Figure \ref{auto} represents the potentiality of these differentiations.
Here, type-0 cells are regarded as stem cells that have the potential 
both to reproduce themselves and to differentiate into other cell types, while 
the differentiated (i.e. type-1, 2, and 3) cells have lost such potential 
and only reproduce the same type.  

In the present example, four distinct cell types are observed.
The number of cell types appearing in the cell society depends on the nature of the network reaction network.

In Fig.\ref{others}, some other examples of the dynamics of distinct cell types and their differentiation rules are displayed, where different reaction networks have been employed.
We see that the number of cell types appearing is often larger than four, as in  Fig.\ref{others}(a),
and the rules of differentiation are often hierarchical, as in Fig.\ref{others}(b).
\footnote{
Of course, existing multicellular organisms (such as mammals) often have hundreds of cell types with more complicated hierarchical rules of differentiation.
The number of cell types that appears in our dynamical differentiation model does not show a clear increase with the number of chemical substances.
We believe that the existence of only a few types of cells is due to the random nature of the chemical reaction network and the fact that the rate constants for all reactions are equal.
In a real biological system, the chemical reaction network is more organized, possibly in a hierarchical manner, through evolution.  
}

\subsection{Diversity of cell colony}

We now consider the existence of multiple cell colony states.
For example, in the automaton rules of the system corresponding to Fig.\ref{others}(b), not all of the six types of cells coexist in a developed cell society.
Cell groups consisting of only two or three cell types appear, depending on the initial condition. 
For example, there is a cell colony  consisting of only type-0, 1, and 2 cells and are consisting of only type-0, 2, and 4 cells.
In this example, we found that four types of cell colonies, i.e., four stable distributions of cell types, appear through developmental processes with different initial conditions (i.e. initial concentrations of chemicals in the first cell), even though the process always starts from a single type-0 cell.
The selection of one cell colony is determined by the distribution of cell types at an early stage of development, when the first differentiations from `0' to `1' and `2' occur stochastically (see Ref.\cite{CFKK1} for details).
It is important to note that each stable distribution of cell types is stable with respect to macroscopic perturbations, including the removal of cells, as is shown in \S \ref{sta_s}.

In our model, depending on the condition of the initial cell, different cell colonies are obtained.
With this result, our model describes a manner in which several types of tissue can appear through the interaction among cells. 
This kind of diversity is often observed in a cultivated system of stem cells. For example, in a hematopoietic system, even if multi-potent stem cells are selected as homogeneously as possible, there is a remarkable variability in the sizes and often in the characters of the developed colonies \cite{Nakahata}.

\section{Robustness of a cell society}
\label{sta_s}

An important implication of our chaos hypothesis for stem cell systems is that a cell society with various cell types that develops according to this hypothesis is stable with respect to certain types of perturbations.
In this section, we discuss this robustness on two different levels at the molecular (microscopic) level and at the  cell-population (macroscopic) level.

\subsection{Stability with respect to microscopic perturbations}

In our model, a cell society with a variety of cell types is stable with respect to molecular fluctuations, even though the process of differentiation is based on the dynamical instability caused by cell-cell interactions.
This cell society is stable with respect to perturbations in the form of an external influence that alters the concentrations of metabolites. 
As long as such a perturbation is not too large, the same set of cell types, a similar distribution of cell types, and similar cell lineages are obtained.   

\begin{figure}[htbp]
\begin{center}
\includegraphics[width=17cm,height=17cm]{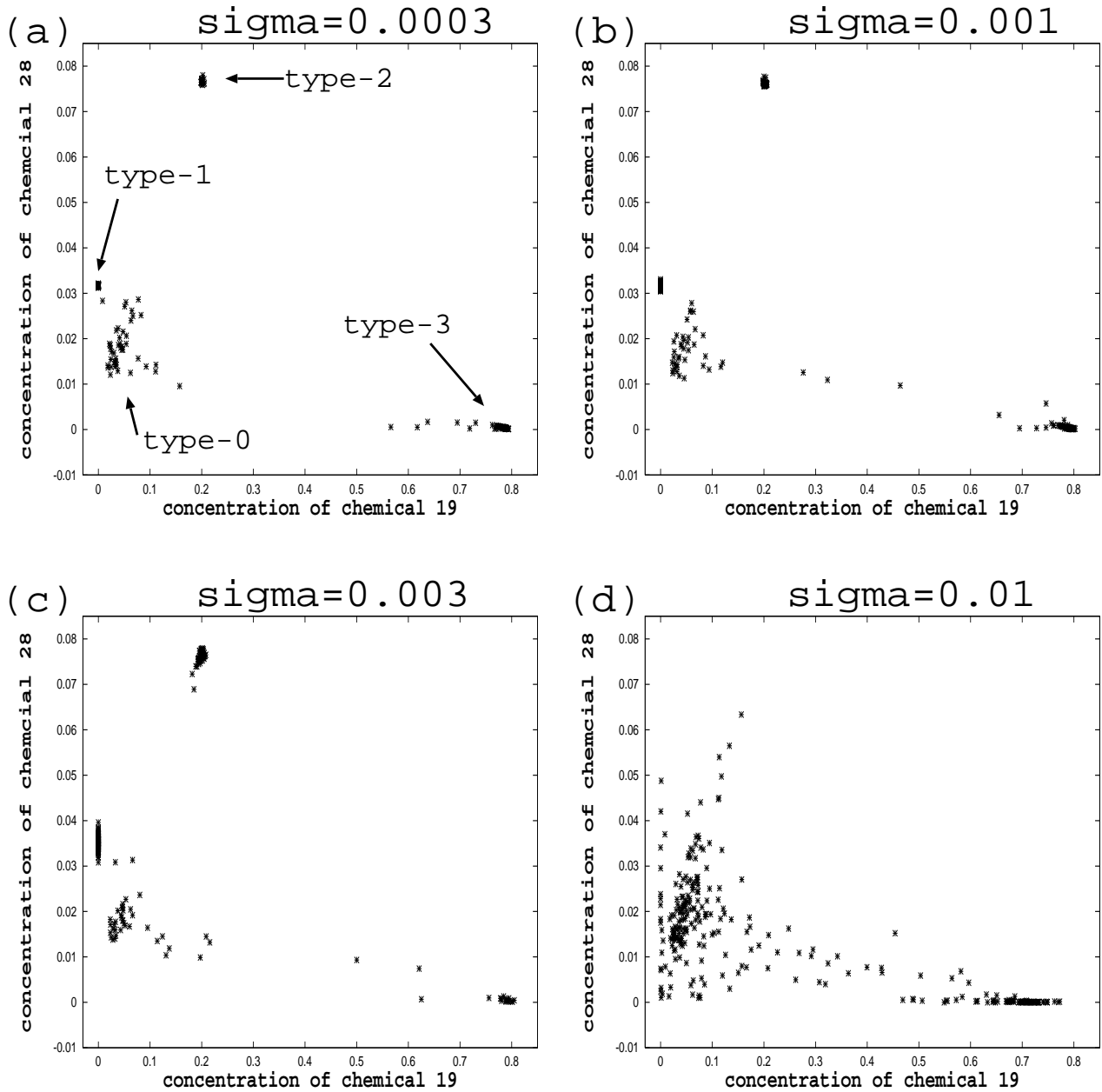}
\end{center}
\caption{\label{noise}
The temporal averages of the concentrations of $x^{(19)}_i(t)$ and $x^{(27)}_i(t)$ for 20,000 time steps.
In the figures, each point represents the state of a cell.
In the situation considered here, the number of cells is 200.
The noise amplitudes $\sigma$ are 0.0003 in (a), 0.001 in (b), 0.003 in (c), and 0.01 in (d).
Some of the points are overlaid and may be invisible since the plotted value of the cells are rather close to each other.
}
\end{figure}

To demonstrate this robustness, we have introduced noise in our simulations to represent molecular fluctuations, described by a Langevin equation.
In Fig.\ref{noise} we have plotted the temporal average of the concentrations of two chemicals ($x^{(19)}$ and $x^{(27)}$) when the number of cells is 200.
In this figure, each point represents the state of a cell.
Without the noise, the cells split into four cell types as shown in Fig.\ref{diff}, which correspond to the distinct clusters of points in Fig.\ref{noise}.
As shown in Fig.\ref{noise}, these four distinct clusters are preserved as long as the amplitude of noise $\sigma$ is less than 0.01.
For larger $\sigma$, the states of differentiated cell types (i.e. type-1, 2, and 3) start to become destabilized, and all cells fall into type-0 dynamics.
It is thus seen that the region occupied by the orbit of type-0 becomes much broader, as the strength of the noise increases (see Fig.\ref{noise}(d)).

This result may be expected when one considers that the differentiated cell types have smaller basins of attraction than the stem-type cell.
Note, however, that the basin of attraction of each cell type depends on the cell-cell interactions.  For example, the measure of the basin of the attraction of type-2 dynamics is zero when a single cell is in the medium, and it increases as the number of type-0 cells increases beyond the threshold.

From several simulations using different reaction networks,  the following common features are found:

\begin{list}{\thectr}{
\usecounter{ctr}
}

\item The dynamics of a stem-type cell generally have a larger basin of attraction than these of differentiated cell types.

\item The nature of the stem-type cell can change significantly due to 
a tiny change in the cell-cell interactions, and has potency to differentiate into other cell types.

\item In spite of instability with respect to change in interactions, 
the stem cell has a larger basin of attraction
in the presence of noise, and its dynamics are stable with respect to noise.

\end{list}

These three features are reminiscent of globally  stable
Milnor attractors, found in a class of coupled dynamical systems \cite{Milnor}.
In such dynamical systems, there exists a certain type of attractor (Milnor attractor) which is easily altered by small structural perturbations
but maintains globally attractive nature, which is often enhanced by noise.
Thus there are examples in the study of dynamical systems of the type of stem-cell dynamics we have been discussing here.

As is discussed in Ref.\cite{KKTY3}, it may be possible to estimate the
minimal number of molecules in a cell required for a robust stem cell system
from the value of the threshold noise $\sigma_{thr}$.
As roughly estimate (see Ref.\cite{KKTY3} for details), the minimal number of molecules necessary for robust development is around 100-1000, which is consistent with observations in experimental biology.

\subsection{Stability with respect to macroscopic perturbations}

In addition to the stability discussed above with respect to molecular fluctuations, our cell society is also stable with respect to against macroscopic perturbations, for example, 
the removal of some cells.

\begin{figure}[htbp]
\begin{center}
\includegraphics[width=9.5cm,height=8cm]{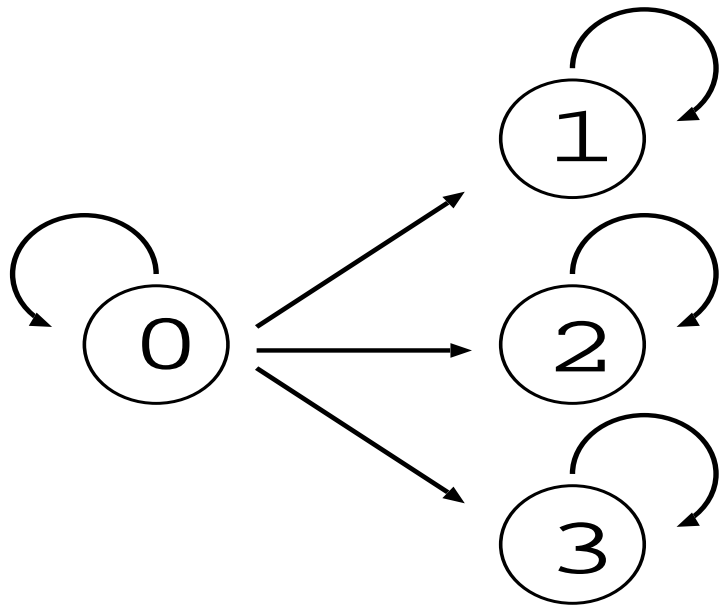}
\end{center}
\caption{\label{auto}
Automaton-like representation of the differentiation rules.
A path returning to the original node represents reproduction the same cell type, 
while paths leading to other nodes represent the potentiality to differentiation 
into the corresponding cell types.
}
\end{figure}

As mentioned above, stochastic differentiation from stem cells obeys 
specific rules, as shown in Fig.\ref{auto}.
When there are multiple choices for a differentiation process (as `0'$\rightarrow$`0', `0'$\rightarrow$`1', `0'$\rightarrow$`2', `0'$\rightarrow$`3' depicted in Fig.\ref{auto}), the probability to select each path is neither fixed nor random, but depends on the number distribution of cell types in the system.
This regulation of differentiation probabilities from stem-type cells into the various differentiated cell types sustain the robustness of a cell society with regard to the populations of each cell type.
As an example, let us consider the case with four cell types (`0',`1',`2',`3' in Fig.\ref{auto}).
When type-1 cells are removed, decreasing their population, the probability of differentiation from a stem cell (type-0) to a differentiated cell of type-1 increase, and the original cell-type distribution is roughly recovered.

\begin{figure}[htbp]
\begin{center}
\includegraphics[width=17cm,height=12cm]{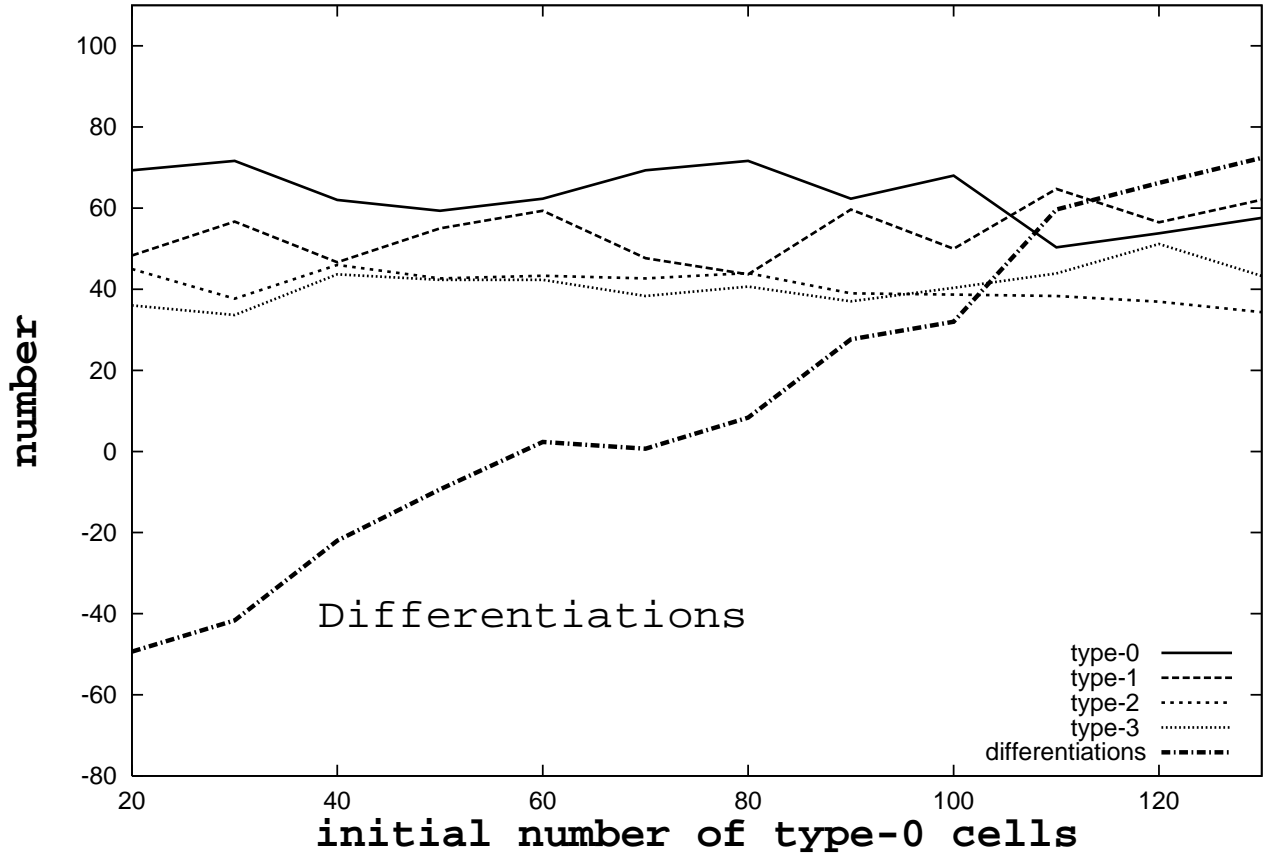}
\end{center}
\caption{\label{sta}
The rate of the differentiation from type-0 to other cell types.
In this simulation, the total cell number was fixed to 200 by disallowing cell division.
We take the initial distribution of the number of cell types as $(n_0,n_1,n_2,n_3)= (n_0,130-n_0,35,35)$, where $n_k$ represents the number of type-$k$ cells.
Starting the simulation with this initial distribution, the number of differentiations from type-0 cells and the final number of each cell type are plotted as functions of the initial number of type-0 cells, $n_0$, when the system settles down to a stable distribution of cell types.
The negative value of the number of differentiations from type-0 cells reflects the occurrence of the de-differentiation process from differentiated cells to type-0 cells, which never occurs in the developmental process from a single cell.
}
\end{figure}

In Fig.\ref{sta}, the rate of differentiation from the type-0 cell to others is plotted, as a function of the number of type-0 cells in the cell ensemble.
In this simulation, to more clearly display the dynamics of the relative number of each cell type, the total number of cells in the medium is fixed (to {\it N}=200 in the present case), by disallowing cell division.
As the initial conditions, $N$ cells are placed in the medium and the 
concentrations of the chemicals in each cell are selected so that they give type-0, 1, 2 or 3 cells.
The switch of cell types were measured when the system settled down to a stable distribution of cell types. 
The simulations ware repeated, changing the initial distribution of 
cell types to obtain the plot in Fig.\ref{sta} of the number of switches from type-0 cells to the others, where the final number of cells for each type is also plotted.  
As shown in Fig.\ref{sta}, the frequency of switches from the type-0 cell increases almost linearly with the initial number of that cell type.
Through these switches, the cell distribution approaches approximately $(n_0,n_1,n_2,n_3)=(60,50,50,40)$, where $n_k$ represents the number of type-$k$ cells.

To maintain the robustness of a cell society, cells must obtain information concerning the distribution of cell types in the system.
According to our result, this information is embedded in the internal dynamics of each cell.
Each attracting state of internal dynamics, corresponding to a distinct cell type,
 is gradually modified in the phase space as the number distribution of the four cell types changes.
However, these 
modifications of internal dynamics are much smaller than the differences among the dynamics of the different cell types.  
Thus, there are two types of information contained in the internal dynamics: ``analogue" information, reflecting the global distribution of cell types, and ``digital" information 
defining the cell type.
This analogue information controls the rate of differentiation, because the probability of differentiation from the stem cell to other cell types depends on the modification of internal dynamics in the stem cell.
On the other hand, a change in the distribution of cell types, resulting from differentiation, 
brings about a change in the analogue information.
As a result of this interplay between the two types of information, a higher-level dynamics emerges, which controls the probability of stochastic differentiation and sustains the robustness of the entire system.

It should be stressed that, in our model, the dynamical differentiation process from stem cells is always accompanied by this kind of regulation process, without any sophisticated programs implemented in advance. 
In this process, differentiations occur when the instability of the system exceeds some threshold through the increase of the cell number, and the emergence of differentiated cells stabilizes the system.
Therefore, a large perturbation, such as the removal of cells, causes the system to become unstable again, but as a result, differentiations that lead the system back toward the stable state occur.
In other words, in the systems that we have studied, only cell types that possess this regulation mechanism to stabilize the coexistence with other cell types can appear through differentiations from the stem cells.

\section{Decrease of complexity in cellular dynamics with the developmental process}
\label{decrease}

In the normal development of cells, there is clear irreversibility,
resulting from the successive loss of multipotency.  ES cells that can create
all types of cells differentiate into several types of stem cells with multi-potency, which produce a
given set of cell types, and then further differentiations from these stem cells follow to 
provide the determined cells, that can produce only the same type of cells. 
How is this irreversibility characterized in our dynamical
systems model of stem cells?

In our model simulations, the dynamics of a stem-type cell exhibit irregular oscillations 
with orbital instability leading to chaos and involve a variety of chemicals.
Stem cells with these complex dynamics have the potential to differentiate into several distinct cell types.
Generally, the differentiated cells always possess simpler cellular dynamics than the stem cells, for example, fixed-point dynamics and regular oscillations (see Fig.\ref{diff} as an example).
The differentiated types of cells with simpler dynamics do not have the ability to differentiate, and they simply reproduce the same type of cells (or stop dividing).
Here, the loss of multipotency
is accompanied by a decrease in the diversity of chemicals and
in the complexity of intra-cellular dynamics. 
Although we have not yet succeeded in formulating the irreversible
loss of multipotency in terms of a single fundamental quantity (analogous to 
thermodynamic entropy), we have heuristically derived a general law describing the 
decrease of two quantities in all of our numerical
experiments, using a variety of reaction networks.
These are discussed below.
\\
~\\

\noindent
{\bf Law of Irreversible Development: Working Hypothesis}

In the normal developmental process starting from a stem cell from which arise various differentiated cell types, the loss of multipotency is accompanied by the following:

\begin{list}{\thectr}{
\usecounter{ctr}
}
\item {\bf A decrease in the diversity of chemicals.}
\item {\bf A decrease in the  degree of orbital instability (`chaos').}
\end{list}

First, we study the diversity of chemicals in the cellular dynamics.
In our model, a variety of chemicals participate in the irregular 
oscillations of the stem-type cell's dynamics, as shown in Fig.\ref{type-0}.
On the other hand, in differentiated cell types, the concentrations of some chemicals fall approximately zero, and the dynamics consist of fluctuations of 
a smaller set of chemicals.
In this sense, the number of dimensions of the
effective phase space of the intra-cellular chemical dynamics become smaller.
To study the decrease of chemical diversity quantitatively, we define 
this diversity for the $i$-th cell by

\begin{equation}
S_i = - \sum_{j=1}^k p_{i}(j) \log p_{i}(j),
\end{equation}
with 
\begin{equation}
p_{i}(j)=<\frac{x^{(j)}_i}{\sum_{m=1}^k x^{(m)}_i}>, 
\end{equation}
where $<..>$ represents the temporal average taken between successive cell divisions.

In Fig.\ref{dire}, we display some examples of the change in the chemical diversity 
accompanying the developmental process.
These correspond to the examples in Fig.\ref{diff} and Fig.\ref{others}(b).
In the figures, each point corresponds to a cell division, and the chemical diversity within that cell, averaged from its most recent division time, is plotted as a function of time.
During the first stage of development ($time<10000$ in Fig.\ref{dire}(a)), cells have not 
yet differentiated, and the dynamics of cells are those with a high diversity of chemicals.
At a later time, differentiated cells begin to appear from the initial cell type.
In the case with hierarchical differentiation depicted in Fig.\ref{others}(b), the diversity of chemicals decreases successively with the development, as shown in Fig.\ref{dire}(b).
These figures clearly show the tendency toward a decrease in the diversity of chemicals along the direction of development.

\begin{figure}[htbp]
\begin{center}
\includegraphics[width=16cm,height=12cm]{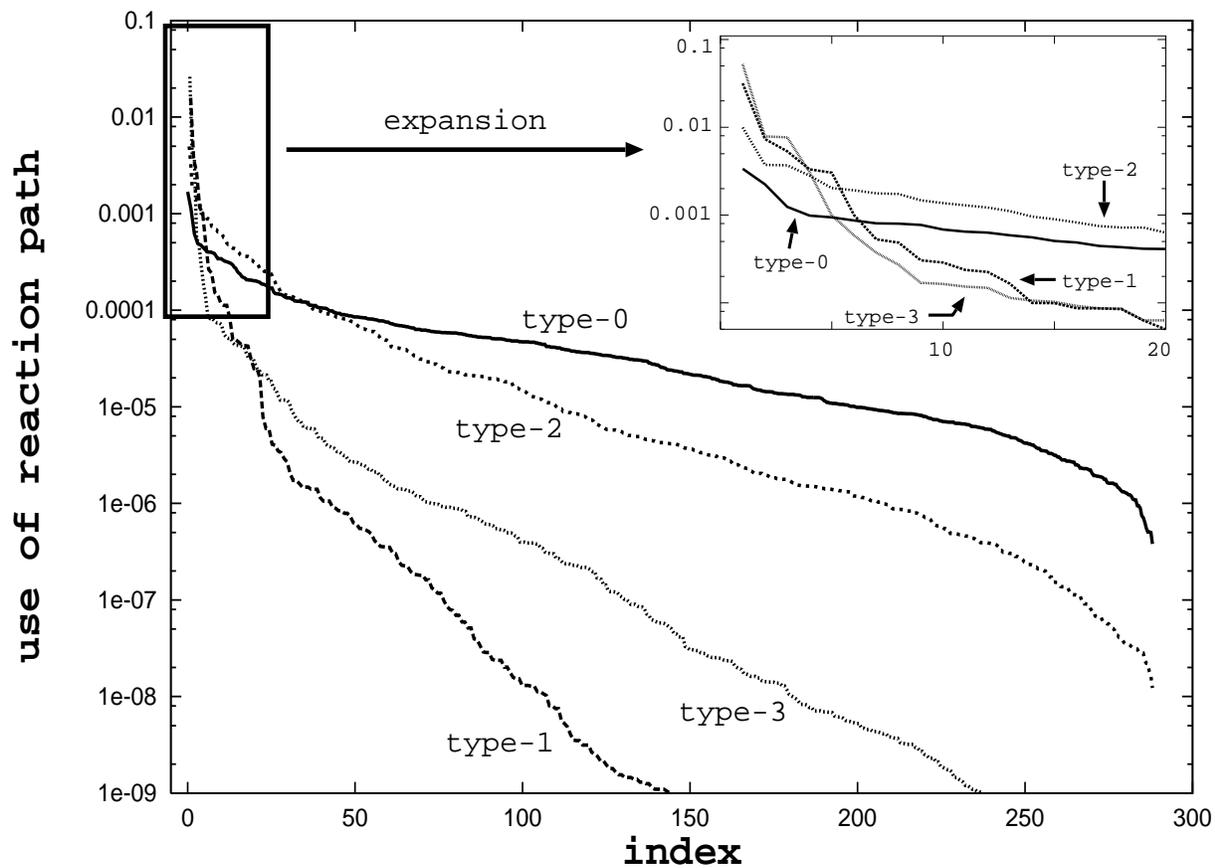}
\end{center}
\caption{\label{path}
Use of reaction paths in dynamics of each cell type.
The amounts of use of each reaction path $m \rightarrow \ell$ 
are computed by the average of the term $Con(m,\ell,j)x^{(m)}_i(t)(x^{(j)}_i(t))^{\alpha}$
over 20000 time steps.  The amount of the use of each reaction path is plotted
in the order of its magnitude, as the horizontal axis representing the
order.
Each line corresponds to each cell type shown in Fig.\ref{diff}.
}
\end{figure}

This decrease in the diversity of chemicals is accompanied by a decrease in the diversity of use of reaction paths in cellular dynamics.
In Fig.\ref{path}, the amounts of use of each reaction path during a certain period are plotted for each cell type.
In this figure, the data are arranged in the order of the magnitude of the use of each reaction path, computed as the average of the term $Con(m,\ell,j)x^{(m)}_i(t)(x^{(j)}_i(t))^{\alpha}$ in eq. (2).
This figure shows that, in type-0 dynamics with complex oscillation and diversity of chemicals, more diverse reaction paths are used than simple dynamics of differentiated cell types, in which a small number of reaction paths are used dominantly while other paths are hardly used.

Next, to confirm the decrease in complexity of the dynamics along with the loss of multipotency, we have measured the `local'
Kolmogorov-Sinai (KS) entropy, which is known to be a good measure of the variety of orbits and the degree of chaos.
Although the KS entropy is defined using all degrees of freedom in the system, including all cells and the medium, here we measure only the complexity of the intra-cellular dynamics by neglecting the instability arising from cell-cell interactions.
We do this by computing the ``local KS entropy'' of the intra-cellular dynamics, defined as the sum of positive Lyapunov exponents, by restricting the set of possible perturbations (the tangent space) to those of only the intra-cellular dynamics of each cell. 
We have found that this local KS entropy of differentiated cells is always smaller than that of stem-type cells before differentiation.
For example, the KS entropy of the stem-type cell in Fig.\ref{diff} is approximately $1.2\times 10^{-3}$, while it is $2.0\times 10^{-5}$ for the type-2 cell, and less than $1.0\times 10^{-6}$ for the type-1 and type-3 cells.
In the case with hierarchical differentiation, the local KS entropy
decreases successively with each differentiation.
These results suggest that the multipotency of a stem-type cell is due to the complexity of its intra-cellular dynamics.

\section{Change in growth speed of cells with the developmental process}
\label{growth_speed}

In our model simulation and in nature, the growth speed of each cell depends
on its cell type and certain characteristics of the cell society environment, such as the number distribution of cell 
types therein.
In our simulations, during the first stage of the development, the stem-type cells
(e.g. the type-0 cells in Fig.\ref{diff}) grow and divide rapidly when the number of cells is small and only 
stem-type cells exist in the system.
However, 
as the number of cells increases and other types of cells appear through differentiation,
the growth speeds of all cells start to depend on their types.
For example, some cell types maintain a high growth speed and undergo successive divisions,
while some others start to lose chemicals into the medium and begin to shrink their volume.

In general, the growth speed of stem-type cells is found to be
smaller than those of differentiated cells, when they coexist in the system.
In Fig.\ref{growth}, we display some examples of the relationship between the growth speed of a type of cell and 
the chemical diversity within the cell, which is a good measure of the degree of differentiation,
as discussed in \S\ref{decrease}.
In the figures, the data points show how the division speed depends on the diversity in the case that there are 200 total cells.
Note that the growth speed is measured by the speed for division in our model.
Each figure corresponds to a different reaction network.
Here,  clusters of points correspond to different cell types, and the cluster with the highest diversity
corresponds stem-type cells.
These examples clearly show the increase of cellular growth speed with the decrease of chemical 
diversity that progresses along the direction of differentiation.
It should be noted that the growth speed of stem-type cells decreases
only when the differentiated cell types begin to coexist with them.
That is, the growth speed of an isolated stem-type cell is not always smaller 
than that of an isolated differentiated cell.

The decrease of the growth speed of stem-type cells
can be explained in terms of a change in the requirement for nutrients.
A stem-type cell with high chemical diversity absorbs a variety of nutrients from the medium.
On the other hand, a differentiated cell type with simple internal dynamics and lower chemical diversity is specialized in the use of only a few kind of nutrients.  
As shown in Fig.\ref{path}, in such cell with simple internal dynamics, a small number of reaction paths are used dominantly and they often form a simple auto-catalytic reaction network.
Since each reaction rate is nonlinear with respect to the involved chemical concentration, the internal reaction dynamics of such simple cell progresses faster.
Then, required nutrients are transformed into non-penetrating chemicals much faster than those in stem-type cell with complex cellular dynamics.
When the stem-type cells and differentiated types of cells coexist, differentiated cells can efficiently obtain the nutrients they require, 
while stem-type cells cannot do the same, or they even may 
release necessary nutrients, because the concentration of these chemicals in the medium is often lower 
than that in the stem-type cells.

Accordingly, the number density of stem cells generally decreases as the total cell number increases.
Let us further consider the case in which the total number does not increase indefinitely, due to
probabilistic deaths of each cell type.  Then, depending on the network structure,
it is expected that either the stem cells continue to constitute a small fraction of the total number of cells and maintain a slow division speed, or the stem cells disappear in the steady state.
In the former case, the spontaneous regulation of differentiation from the stem cells
insures the stability of the cell number distribution with respect to changes in the 
death rates of
each cell type, while in the latter case, the cell number distribution of differentiated 
cell types 
is vulnerable to such changes.

\section{Discussion and Summary}
\label{Discussion}

To confirm the validity of our chaos-based model of stem-cell systems, we have carried out numerical experiments using several different sets of parameter values and 
choosing thousands of different reaction networks.
As a result, we have found that the scenario discussed in preceding sections 
is commonly observed for some fraction of randomly generated reaction networks.
With the parameter values used in the example considered in 
Fig.\ref{diff}, approximately 10\% of randomly chosen reaction networks result in oscillatory behavior, while others converge to fixed points.
Furthermore, approximately 20\% of these oscillatory dynamics are destabilized through cell division.
In these cases, the diversity of cell types emerges as discussed above.

One may ask why we select such complex oscillatory dynamics to describe general mechanisms for the dynamics of stem cells, while only a
small fraction of randomly chosen reaction networks lead to such oscillatory dynamics.
In anticipation of this question, we have studied the growth speed of
an ensemble of cells to determine
what kind of reaction networks can possibly appear through evolution.
As is shown in Ref.\cite{PRL}, if the internal dynamics
can be chaotic and if the cell ensemble contains a variety of cell types differentiated from stem-type cells, the cell society can attain
a larger growth speed as an ensemble by realizing a cooperative use of 
resources, in comparison with an ensemble with homogeneous cells and simpler internal dynamics.
This suggests that the emergence of a stem-cell system with complex cellular dynamics and a diversity of cell types is a necessary course in the evolution of multicellular organisms, as aggregates of cells compete for finite resources necessary for growth and reproduction.
From this reasoning, we find that the use of reaction networks leading to 
complex oscillatory dynamics for the description of stem cells is also supported by evolutionary consideration.

Note that the results found in this paper are changed little when we change the model parameters.
They are essentially unchanged as long as the magnitudes of the internal reaction term and the cell-cell interaction term are of comparable order, and the set of equations describing the internal reaction are sufficiently complex to exhibit non-linear oscillatory dynamics that in some case can be chaotic.
Such complex internal dynamics are observed with some fraction of 
randomly generated reaction networks,
when the number of chemicals is sufficiently large (say for $k>10$) and the number of the reaction paths is in the intermediate range.
For example, when $k=32$, there must be approximately 6-12 paths for each chemicals.
If there are fewer or more. the intra-cellular dynamics converge to a fixed point for most cases, without differentiation.
The results are also independent of other details of the model.
For example, they are unchanged if we change the catalyzation degree $\alpha$ to 1 or 3.

Let us summarize our results and their implications.

First, as a cell with initially oscillatory dynamics and a variety of chemical species replicates, the divided cells possess chemical compositions similar to that of their mother cell.
Then, when the cell number increases beyond some threshold value, some of the cells start to differentiate into different types, that can clearly be distinguished by different (average) chemical compositions.
In terms of dynamical systems, each cell type corresponds to a pattern of orbits in a different location in the chemical phase space.
The differentiated state in a given cell can persist and be transmitted to the daughter cell, or it can switch to a different state, which is then transmitted to the daughter, according to a particular set of possible pathways allowed by differentiation rules.
These rules are often hierarchical, as in Fig.\ref{others}(b).  

Cell differentiation from stem cells
and the rules governing it are not introduced explicitly into the model, but, rather, 
they emerge through the instability of the intra-cellular dynamics induced by cell-cell interactions.
Due to this dynamical instability, the behavior of a stem cell with regard to reproducing its own type or differentiate into a new type is stochastic.
This behavior cannot be determined only from the extra-cellular conditions, such as the concentrations of signal molecules.

Even though the differentiation process results from a dynamical instability, the developmental process and the states of the various cell types are stable with respect to microscopic perturbations, such as fluctuations at the molecular level, and with respect to macroscopic perturbations, for example, the removal of cells.
An important point is that this robustness of the system is a natural consequence of this cellular diversification process, and it does not
require any sophisticated mechanisms.
The differentiation process in our model occurs when the instability of the whole system exceeds some threshold, and the resulting coexistence of several cell types re-stabilizes the overall cellular dynamics.
Here, only the cellular states well separated in phase space that are stable with respect to microscopic perturbations can appear through the differentiations from the stem cells, and only cell type distributions that are stable with respect to macroscopic perturbations are realized.

Along with the developmental process in the cell society, from multipotent stem cells to differentiated cells, we have found that there is a decrease in the complexity of the intra-cellular dynamics.
Quantitatively, this irreversible change accompanying the developmental process
is represented by a decrease of the chemical diversity, the diversity of use of reaction paths, and KS-entropy of the 
intra-cellular dynamics.
Additionally, when stem cells and differentiated cells coexist, the growth speed of stem cells is generally smaller than that of differentiated cells.

It should be stressed that, it is not the aim of the 
present study to fit the results of simulations to experimental data, such as the ratio of stem cells in colonies of hematopoietic cells
\footnote{
In our model simulations, due to the stochastic behavior of stem cells, the distribution 
of the number of stem cells in each colony is often broad, as observed in the hematopoietic system \cite{Till}, when simulations start from different initial conditions (i.e. different initial chemical
concentrations).
In other cases with different reaction networks and parameters, this distribution of the number of stem cells has a sharp peak; that is, the number of stem cells is almost the same
for each colony.
Additionally, we have found that the distribution has a sharper peak when the spatial configuration is taken into account in the model, 
for example, by placing the cells in a 2-dimensional medium so that they interact with each other through the diffusion of chemicals in the medium \cite{CFKK2}.}.
Rather,  it is our goal to realize the essence of stem cell systems, 
including spontaneous stochastic differentiation, the robustness of the system, and the 
irreversible loss of multipotency, as a general consequence
of a cell society with dynamical complexity.
In this study, we have given evidence suggesting that the characteristics of a stem-cell system
are a universal feature in a system with complex internal dynamics,
interactions and duplication.

In the study presented here, spatial variation among cells, which plays an essential role for morphogenesis, is not taken into account.
By introducing cell-cell interactions through the diffusion of  chemicals in  
 2 dimensions, we have found that the differentiation from stem cells,
as described above, is regulated spatially.
This regulation leads to an ordered spatial pattern of differentiated cells \cite{CFKK2}, as seen in apical meristem and interstitial crypts.
This spatial pattern is generally stable with respect to external perturbations, 
for example, the removal of cells.  After such damage has been sustained by the system, 
stem cells differentiate in such a way to restore the original pattern, and
the damaged part is thus repaired.
An important point here is that in the present model the spatial information used to control a cell's fate, such as the gradient of signal molecules, is not supplied from outside of the system, as is often done \cite{cell}.
Instead, this spatial information emerges through the cell-cell interactions in such a way to sustain the robust cell society.

\section{Predictions}

We believe that our results are universal in a class of dynamical systems
that extract the minimal factors in the developmental process,
ant it is thus expected that the real stem-cell systems possess 
the same features.  To close our paper, we
summarize the predictions we can make using our model, and discuss the possibility
of experimental verification.

The main claims regarding stem systems we make based on our model simulations are as follows.

\begin{enumerate}

\item
The chemical reaction dynamics in a stem cell are of a complex oscillatory 
nature with a chaos-type instability.
These dynamics are responsible for the stochastic behavior of stem cell differentiation.

\item
Stochastic differentiation from a stem cell is induced by the cell-cell interactions among surrounding cells.
Note also that differentiation is not necessarily 
simultaneous with cell division.

\item 
  The probability for a stem cell to reproduce its own type of cell or to differentiate into a different type of cell depends on the environment, i.e., the distribution of other cell types.
This regulation of the differentiation probability leads to the stability of the cell society 
with respect to external perturbations.

\item
  The irreversible loss of multipotency experienced in the change from a stem cell to a differentiated cell
is characterized by (i) a decrease in the chemical diversity in a cell and (ii) a decrease of
the (chaotic) instability in the dynamic change of chemical concentrations.

\item
   In a colony consisting of stem cells and other proliferating differentiated cell types, the former has a lower growth speed.  In general, there is a 
negative correlation between the growth speed and the chemical diversity 
in cells when they coexist in a colony.

\item
The growth speed of a colony is an increasing function of the diversity of its cell types.

\end{enumerate}

These assertions provide predictions that can be experimentally verified.   
We now discuss each of them.\\

\noindent
{\bf Chaotic reaction dynamics in a stem cell as a source of stochastic behavior}

In our theory, the stochastic behavior of stem cell differentiation results from non-linear, chaotic chemical reaction dynamics existing within the stem cell.
Although such complexity of intra-cellular dynamics has not yet been observed, we believe that 
they are essential in the behavior of stem cells.
It was recently reported that
the  $Ca^{2+}$ oscillation in a cell depends on the cellular state, and
that can change the pattern of gene expressions \cite{Ca}.
One possible way to check our conjecture is simply to study the oscillatory reaction dynamics, such as $Ca^{2+}$ oscillation, in a cell and check how the irregularity in the oscillation
differs between a stem cell and a differentiated cell.
Studying the dynamic change of gene expression patterns is also a 
possible way to check our theory.
\\

Plasticity in the states of stem cells has been found experimentally.
For example, neural stem cells were found to be capable of re-constituting the hematopoietic system in irradiated adult mice \cite{plast}.
These observations suggest that the behavior of stem cells can change
flexibly according to the environment, while differentiated cells have
no such flexibility.  In our theory also, the variability of cellular
behavior decreases from a stem to differentiated cells.
Indeed, in our numerical experiments, when
the concentrations of nutrients are changed even slightly in the medium, stem cells often undergo significant variation, and as a result new types of differentiated
cells can appear.
In our theory,
this flexibility of stem cells results from the complexity of their internal dynamics with a variety of chemical species, and we expect that this is true also in real stem-cell systems.
~\\

\noindent
{\bf Interaction-based differentiation}

In our theory, the differentiation of a stem cell is induced by interactions between cells in the system.
Thus, we believe that in real systems, the behavior of each cell in a stem cell system depends 
strongly on the existence of and the states of surrounding cells.
For example, we expect that if a stem cell is maintained in an isolated environment by 
continuously removing divided cells, it will not be able to 
differentiate and will only replicate itself.

Recall that in our model, differentiation of a stem cell is not necessarily 
concurrent with its cell division.
Rather, this differentiation occurs when the instability of the cellular dynamics
exceeds some threshold, and this is caused by the increase of cell number through 
cell division.
We believe that similar behavior can be found in real systems, in particular that the 
differentiation rate of a stem cell can be varied by changing the distribution of other cell types and that this differentiation is not necessarily coincident with cell division.\\
\\

\noindent
{\bf Regulation of the differentiation probability leading to the robustness of the cell society}

As mentioned in previous sections, the probabilities governing differentiation of
a stem cell are not fixed in our theory, but are controlled by the distribution of surrounding cell types.
An important point here is that this regulation of differentiation occurs in such a manner that
the effect of an external perturbation is compensated, in analogy to the Le Chatlier principle in
thermodynamics.  
This regulation is that which creates the robustness of the system.

By controlling the cell-type distribution, one can examine experimentally if the rate of differentiation of stem cells is regulated so as to compensate for external change.
For example, if the death rate of one type of cell is increased, or cells of this type are continually removed, 
the differentiation rate from the stem cell into that type of cell should be increased.

In addition, a change in cellular dynamics may regulate the probabilities of stem cell differentiation.
For example, we expect that a change in the dynamics of $Ca^{2+}$ oscillation, which is believed to play an important role in regulating cellular dynamics, results in a change in differentiation probabilities.\\
\\

\noindent
{\bf Irreversible loss of multipotency characterized by decrease of complexity in cellular dynamics}

In our numerical experiments, stem cells, which have the potential to differentiate, possess complex cellular dynamics 
with irregular oscillations and a high diversity of chemical species, while differentiated cell types always possess 
simpler dynamics, for example, those  with fixed-points or regular oscillations, with a low diversity of chemical species.
We have characterized this change of cellular dynamics by the decrease of chemical diversity, decrease of diversity in use of reaction paths, and the decrease of instability of the dynamics, measured by the KS entropy.
The results of the present study suggest that, this change in dynamics along with development is 
universal in systems composed of cells with 
intra-cellular dynamics and cell-cell interactions among these cells.

In terms of molecular biology, the decrease in diversity of chemicals and diversity of use of reaction paths in cellular dynamics can 
be interpreted as the decrease of the number of expressed genes.
This decrease of the number of expressed genes, indeed, has been observed in real systems.
For example, multipotent progenitor cells in hematopoietic systems are found to co-express 
lineage-specific genes of several differentiated cell types, such as erythroid and myeloid cells \cite{Hemato_diversity}.
It should be noted that the levels of this co-expression in multipotent cells are substantially lower than 
those seen in mature lineage-committed cells.
It is suggested by experimental findings that this low level co-expression in the multipotent cell is sustained by complex 
reaction networks including positive- and negative-feedback reactions.
Here, our conjecture in this regard is supported by existing experimental observations.
Considering the generality of our conjecture, it is relevant to check how the diversity of gene expressions
changes in the course of development from a stem cell.
Furthermore, we conjecture that not only the `static' diversity but also
the `dynamic' complexity in intra-cellular dynamics decreases along with the
developmental process from a stem cell.  Again, the measurement of the oscillation pattern of the concentrations of some chemicals (say $Ca^{2+}$) is a possible method to check this conjecture.

While during the normal course of development, this loss of multipotency is irreversible, it is 
possible to recover the multipotency of a differentiated cell through perturbation, by changing
the diversity of chemicals or the complexity of the dynamics.  For example, by expressing a variety of genes compulsively in differentiated cells, the original multipotency may be regained.
Note that, according to our model simulations, the basin of attraction of the stem cell is much larger than that of differentiated cells.
This implies that by adding a large perturbation that results in the presence of a variety of chemicals in a cell, the cell de-differentiates back into a stem cell.
\\

\noindent
{\bf Smaller growth speed of stem cells than differentiated cells}

As mentioned in \S \ref{growth_speed}, in our simulation the growth speed of stem cells is generally smaller than that of differentiated cells, when the stem and differentiated cells coexist.
In many cases, the stem cells stop growing or begin to shrink after the number of
cells differentiated from them has increased to a certain level.
In general, there is negative correlation between the chemical diversity (or the diversity in
gene expressions) and the growth speed, which can be experimentally examined.

Of course, in a real biological system, the activation or suppression of cell growth and division involves
several factors, including signal molecules called ``growth factors", in addition to the condition for nutrients.
Still, we believe that this smaller growth speed of stem cells is a general characteristic of the diversification 
process in stem cell systems.
In our numerical simulations, the speed of differentiated types is greater because they are specialized in obtaining certain 
specific nutrients, in contrast with the stem cells, 
which must obtain a larger variety of nutrients.
We suspect that this may reflect the situation in real systems in which 
stem cells require diverse growth factors for their growth,
while differentiated cells are more specialized in some specific factors.
    
In real multicellular systems, stem cells, such as, neural or hematopoietic stem cells,
often enter a quiescent state (i.e. a $G_0$ state) without division.
These stem cells resume division actively when some external condition is changed, for example, 
an injury that decreases the population of differentiated cells.
On the other hand, differentiated cells, such as lineage-committed progenitor cells 
in the hematopoietic system and interstitial crypt, generally have a faster growth speed.
Here, as an experimentally verifiable proposition, 
we conjecture that the growth of stem cells generally stops if they are surrounded by a 
sufficient number of 
differentiated cells (that are produced by that stem cell), while the decrease of the number of surrounding cells 
leads to a resumption of the division and differentiation.
\\

\noindent
{\bf Positive correlation between growth speed of cell ensemble and cellular diversity}

As shown in a previous work \cite{PRL}, an ensemble of cells with a variety of cell types that is sustained by 
differentiations from stem-type cells generally has a larger growth speed than an ensemble of homogeneous cells, 
because of the greater capability of the former to transport and share nutritive chemicals.
This {\sl positive} correlation between the growth speed and the diversity of cell types (at the cell ensemble level)
can be experimentally verified in any stem-cell system.
For example, if a stem cell loses the ability to differentiate through mutation, 
tissue consisting only of the homogeneous cells has smaller growth speed as a whole
than tissue of the wild-type with various cell types.
In general, it is believed that tissue with a smaller number of cell types
has a lower growth speed as an ensemble of cells.
In fact, this correlation is suggested by studies of cultures of the hematopoietic system \cite{Suda}.
Further study of this correlation will also be
important to understand the evolutionary process of multicellular organisms.
\\\\
{\sl acknowledgements}
\\\\
The authors are grateful to  T. Yomo, T. Ikegami, and S. Sasa 
for stimulating discussions.
The work is supported by 
Grant-in-Aids for Scientific Research from
the Ministry of Education, Science and Culture of Japan
(11CE2006;Komaba Complex Systems Life Project; and 11837004).
One of the authors(CF) is supported by research fellowship from Japan Society
for Promotion of Science.

\clearpage

\begin{figure}[htbp]
\begin{center}
\includegraphics[width=13cm,height=22cm]{./fig5.eps}
\end{center}
\caption{\label{others}
Some other examples of dynamics of distinct cell types and rules of differentiation.
Each figure is obtained by adopting a different reaction network, generated randomly.
In each node, the time series of chemical concentrations in a cell are shown as in Fig.\ref{diff}, representing the course of differentiation to the corresponding cell type.
Each arrow between the nodes represents the potentiality to differentiation to the corresponding cell type, as in Fig.\ref{auto}.
}
\end{figure}

\begin{figure}[htbp]
\begin{center}
\includegraphics[width=11cm,height=22cm]{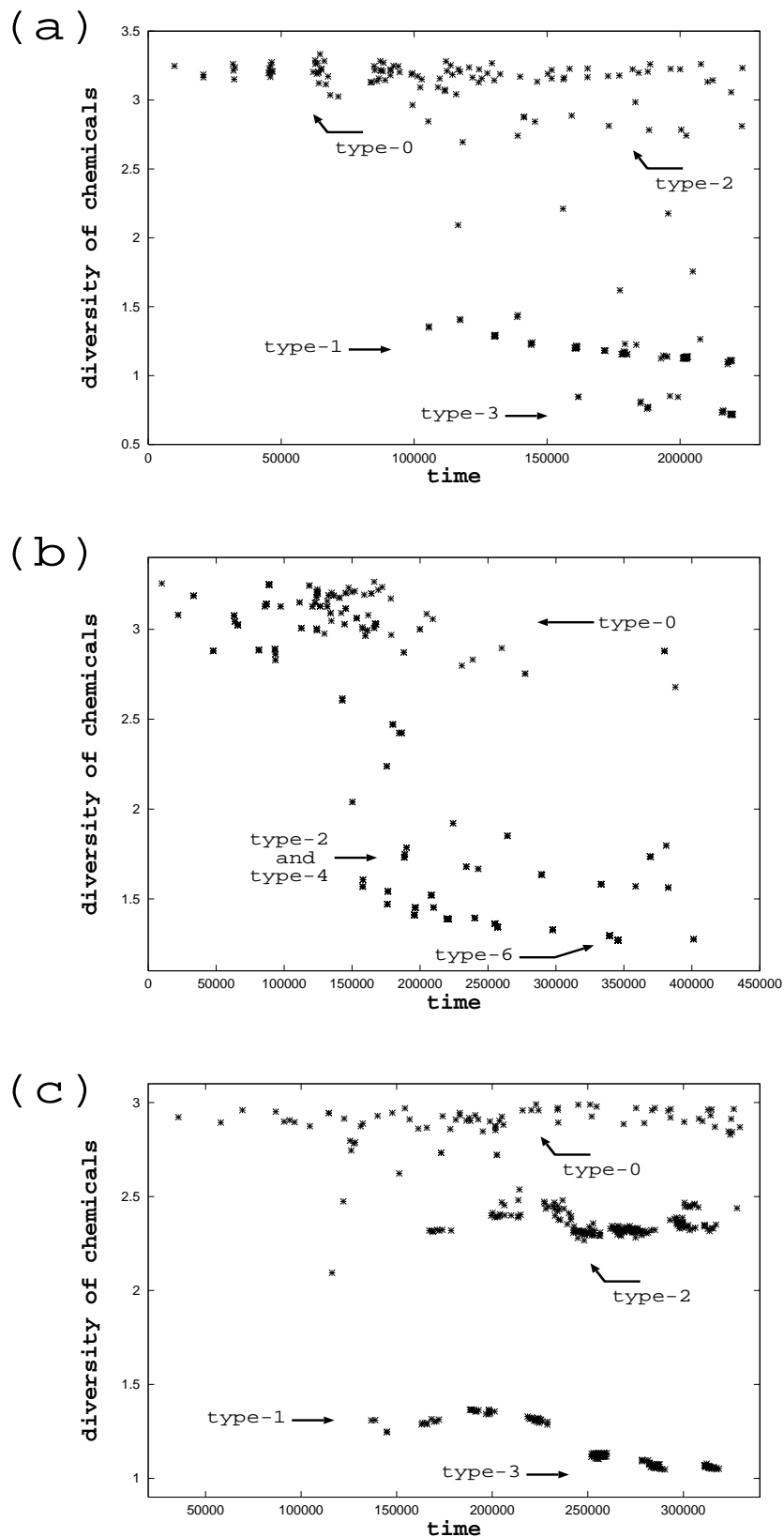}
\end{center}
\caption{\label{dire}
Examples of the change in the diversity of chemicals through the developmental process.
In each figure, the vertical axis represents the diversity of chemicals defined by Eq.(5), while the horizontal axis represents the time.
Each point gives the average chemical diversity of a cell at a cell division, where the average is taken from the most recent division time.
The figures (a),(b), and (c) display the results obtained by using the examples in Figs.\ref{diff}, \ref{others}(a), \ref{others}(b), respectively.
}
\end{figure}

\begin{figure}[htbp]
\begin{center}
\includegraphics[width=12cm,height=22cm]{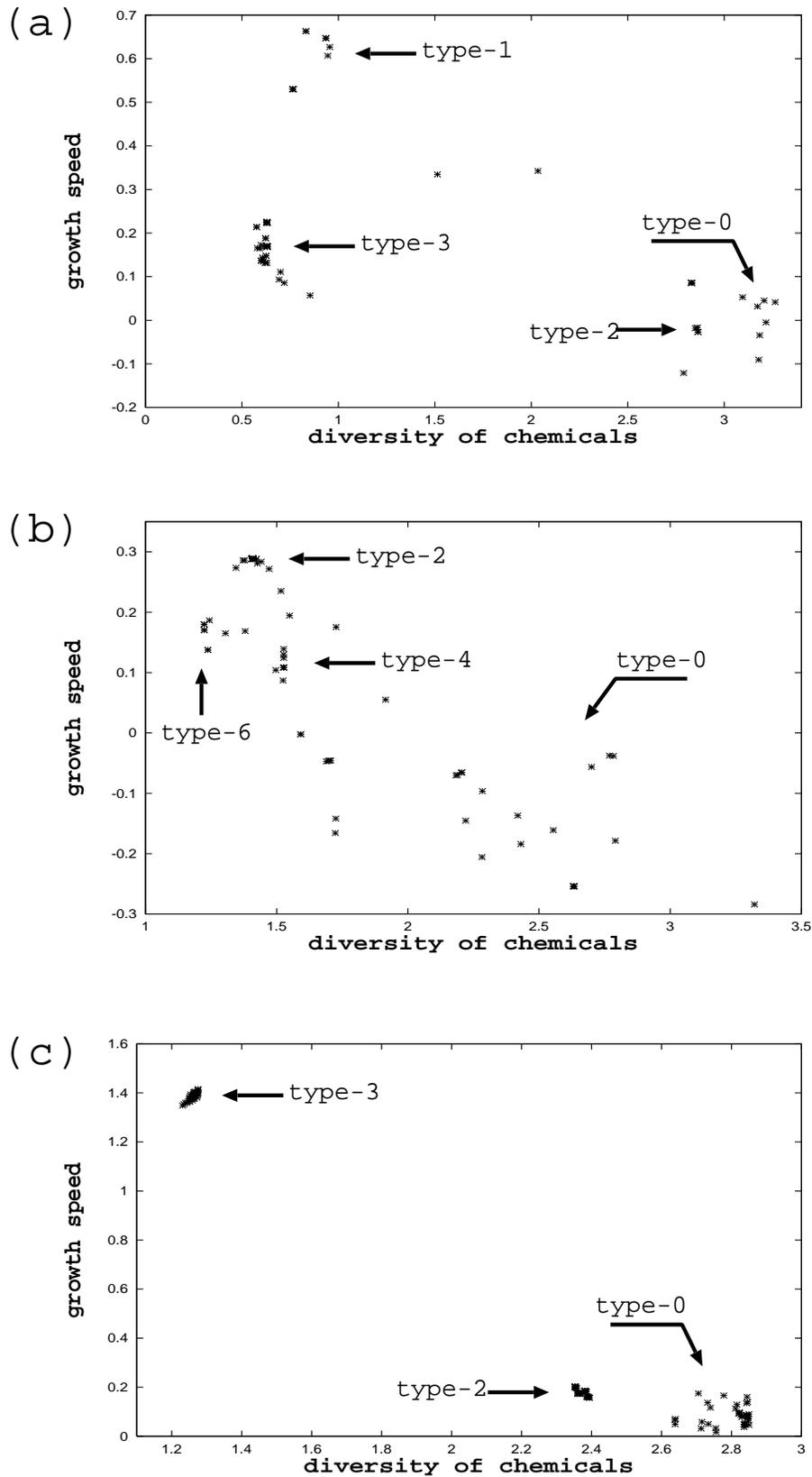}
\end{center}
\caption{\label{growth}
Relation between the division speed and the chemical diversity within a cell.
The figures (a),(b), and (c) display the results obtained by using the examples in Figs.\ref{diff}, \ref{others}(a), \ref{others}(b), respectively.
In the figures, the data points show how the division speed depends on the diversity in the case that there are 200 total cells.
Some of the points are overlaid and may be invisible since the plotted value of the cells are rather close to each other.
}
\end{figure}

\end{document}